\documentclass[12pt, draftclsnofoot, onecolumn]{IEEEtran}
\IEEEoverridecommandlockouts
\usepackage{verbatim}
\usepackage[utf8x]{inputenc}
\usepackage{epstopdf}
\usepackage{graphicx}
\usepackage{tabularx}
\usepackage{booktabs}
\usepackage[table]{xcolor}
\usepackage{booktabs}
\usepackage{multirow}
\usepackage{ctable}
\usepackage[cmex10]{amsmath}
\usepackage{algorithm}
\usepackage{algpseudocode}
\usepackage{pifont}
\usepackage{amssymb}
\usepackage{cite}
\usepackage{subcaption}
\usepackage{textcomp}
\usepackage{threeparttable}
\usepackage{hhline}
\usepackage{array}

\newcommand*\xbar[1]{%
	\hbox{%
		\vbox{%
			\hrule height 0.5pt 
			\kern0.5ex
			\hbox{%
				\kern-0.1em
				\ensuremath{#1}%
				\kern-0.1em
			}%
		}%
	}%
}

\begin{document}
\bstctlcite{IEEEexample:BSTcontrol}
\title{Unlicensed Spectrum Sharing for Massive Internet-of-Things Communications} 
\author{Ghaith Hattab and Danijela Cabric
\thanks{Ghaith Hattab and Danijela Cabric are with the Department of Electrical and Computer Engineering, University of California, Los Angeles, CA 90095-1594 USA.}}  

\maketitle

\begin{abstract}
The unlicensed spectrum, although free, has become an invaluable resource toward enabling massive Internet-of-things (IoT) applications, where Internet-enabled devices are deployed at a large scale. However, realizing massive IoT connectivity over unlicensed bands requires efficient spectrum sharing among IoT devices and fair coexistence with other wireless networks. In this article, we discuss several spectrum sharing methods to address these intra- and inter-network sharing issues. To this end, we first consider ALOHA-based networks that avoid spectrum sensing, yet rely on diversity to improve connection density of random access. Then, we present sensing-based solutions such as unlicensed cellular access that help support a wider range of applications with different rate requirements and connection densities. Finally, we highlight future research directions for massive IoT connectivity over the unlicensed spectrum.
\end{abstract} 
\vspace{0.5in}

\section{Introduction}
The emergence of the massive Internet-of-things (IoT) market, i.e., applications with large-scale deployment of Internet-enabled sensors or \emph{things}, is transforming many vertical sectors \cite{MGI2018}. For example, few cities are already integrating IoT devices with street lights to reduce their power consumption, with traffic signals to improve transportation, and with buildings to monitor their health \cite{Mehmood2017}.

While different massive IoT services can have distinct communications requirements, all of them need wide-area connectivity. Nevertheless, the traffic characteristics of massive IoT set them apart from human-type traffic, requiring wholesale changes to traditional wireless networks. For example, massive IoT traffic is predominately uplink and delay-tolerant.  In addition, existing networks are optimized to maximize the network capacity, whereas the deployment of  battery-powered IoT devices, some of which are installed deep indoors, centralizes the design of IoT networks around coverage, energy-efficiency, and connection density. For instance, the IMT-2020 vision identifies massive IoT as a use-case for next-generation networks, with a connection density of $10^6$/km$^2$  and device lifetime beyond 10 years \cite{ITUR2015}.

The aforementioned properties of massive IoT communications have spurred mobile network operators (MNOs), technology vendors, and standard bodies, to develop tailored wide-area networks in both licensed and unlicensed bands. Although the use of a licensed spectrum ensures strict quality-of-service guarantees, it is limited and expensive to own. Thus, unlicensed bands remain invaluable resource toward enabling massive IoT. Indeed, a new class of low-power wide-area (LPWA) networks has emerged, consisting of various proprietary technologies. The two most prevalent ones are LoRa, that uses chirp spread spectrum (CSS), and Sigfox that relies on frequency hopping (FH) and ultra-narrowband (UNB) signals \cite{Raza2017}. Perhaps less recognized is also the interest of MNOs to capitalize on the wider bandwidth of the unlicensed spectrum by developing cellular standards with unlicensed access, e.g., licensed-assisted access (LAA) and MulteFire, which can carry out private IoT networks \cite{Labib2017,MulteFire2017}. 

The unlicensed-based solutions for massive IoT aim to address two key issues: the \emph{intra-network sharing}, i.e., how a large density of low-cost IoT devices should share the spectrum, and the \emph{inter-network sharing}, i.e., the coexistence of IoT networks with other incumbent networks. In this article, we focus on the technical solutions that address these spectrum sharing issues. To this end, we first summarize the key regulations of different unlicensed bands and their effects on the design of IoT networks. Then, we consider ALOHA-based and sensing-based IoT networks. In the former, access is random, where devices use the spectrum without coordination. In the latter, sensing is first performed, and the identified available resources are then coordinated between the base station (BS) and its IoT devices.

We remark that other works have studied random and coordinated access \cite{Zucchetto2017,Zhang2018a}. However, in this article, we focus on the spectrum sharing aspect, delving into the main access techniques used for massive IoT connectivity. For example, we highlight the role of diversity schemes inherited in ALOHA-based access and present additional enhancements to improve coverage and connection density. We then overview sensing-based networks, outlining the limitations of existing methods and the need to revisit sensing for massive IoT applications. We further discuss a low-cost distributed wideband sensing architecture that enables such networks to support a wider range of applications with different rate requirements and connection densities. Finally, we present some future research directions for massive IoT connectivity over the unlicensed bands.

\section{Properties and regulations of the unlicensed bands}\label{sec:deployment}
There are many candidate unlicensed bands for massive IoT, yet each one has specific regulations to follow, which can also vary by the country or region. The major differences among these bands typically include:
\begin{itemize}
	\item \textbf{The duty cycle:} The maximum duration for an active transmission during a one-hour period 
	\item \textbf{Transmit power:} The maximum transmit power or the maximum equivalent isotropically radiated power (EIRP)
	\item \textbf{Polite spectrum access:} Whether sensing, e.g., listen-before-talk (LBT) or dynamic frequency selection (DFS), is required. 
\end{itemize}
We also note that there are other regulations on the spectral mask used by the device, the support of power control, the type of modulation used, and restrictions to indoor deployments.  
The aforementioned differences have direct influence on the characteristics of massive IoT networks, and so we discuss the key regulations and properties of three main bands: the sub-1GHz, the 2.4GHz, and the 5GHz bands.	

\subsection{The sub-1 GHz bands}
The sub-1GHz spectrum is primarily used by LPWA networks and WiFi HaLow (IEEE 802.11ah) because of its favorable propagation conditions, i.e., lower path and penetration losses. This helps the network deploy fewer BSs when covering a wide area, compared to networks using higher bands. Nevertheless, the sub-1GHz is not universal, and the spectrum available is rather limited. For instance, in the US (or Europe), the band comprises 26MHz (or 7MHz) of the spectrum at 902-928MHz (or 863-870MHz). 

Because polite spectrum sensing is not mandated in sub-1GHz bands in Europe, there are stricter regulations on transmit power and duty cycles. For example, the subband 868.0-868.6MHz, which is used by LoRa and Sigfox, has a peak radiated power of 14dBm and a one percent duty cycle, i.e., an IoT device can transmit only for a duration of 36s per hour. Duty cycles have put strict limits on the number of daily packets the Sigfox network supports per device. Indeed, Sigfox's UL data rate in Europe is 100bps, and each single packet is 208bits, which is sent three times at different frequencies, limiting the number of packets per day to 144. We note that the use of LBT extends the duty cycle to 2.8 percent instead of one percent, and the other subbands in 863-870MHz can have other requirements.

In the US,  there are no restrictions on the duty cycle over the 915MHz band (902-928MHz). However, the dwell time, i.e., the time on air of a single transmission over a particular channel, is enforced for FH systems, which can be an issue for low-rate FH IoT networks. In particular, FH systems, with hopping channels being less than 250KHz, need to use at least 50 hopping frequencies, while occupying each one for at most 0.4s within a 20s period. We note that the FH device does not have to use all the hopping frequencies during each transmission. Since Sigfox's packet takes 2s to transmit 208bits at 100bps, Sigfox has increased the data rate to 600bps using 600Hz channels for the US specifications. Further, the peak power for FH and CSS systems is 30dBm, which is significantly higher than that in Europe. In other regions such as Asia Pacific, different countries have varying requirements, primarily for bands around 433MHz,  779-787MHz, 860MHz, and 920MHz.

\subsection{The 2.4GHz bands}
Different from the fragmented sub-1GHz bands worldwide, the 2.4GHz band is global, enabling economies of scale. In addition, there are more channels spanning the 2402-2484MHz spectrum, and the peak power is typically 20-30dBm, depending on the region. In addition, there are no regulations on the duty cycle, yet in the US, the dwell time can be limited, whereas polite spectrum access may be further required in Europe.

The key issue of this band is the presence of several technologies such as WiFi, Bluetooth, and Zigbee. Nevertheless, there are few IoT networks that use this spectrum such as Ingenu, which uses a proprietary scheme known as Random Phase Multiple Access in order to be robust to interference in this crowded spectrum.

\subsection{The 5GHz bands}
The wide available spectrum at 5GHz, which is approximately 500MHz, has motivated the development of 3GPP cellular standards with unlicensed access \cite{Labib2017}. While the main driver for unlicensed-based cellular networks has been boosting the network capacity, the 5GHz spectrum is now envisioned to also support private IoT networks. For example, the MulteFire network aims to provide IoT-based services by connecting LTE-M and NB-IoT devices over unlicensed carriers, without anchored licensed channels \cite{MulteFire2017}. 

This spectrum is divided into several bands, and the peak transmit power in each one ranges from 23dBm to 30dBm. In addition, LBT is mandated at 5GHz in Europe and Japan, whereas DFS is required in some bands in the US. For this reason, cellular-based solutions use sensing-based access. For example, in LBT, if the channel's measured energy is larger than a threshold, then the BS initiates a random backoff or defers for an extended period. In DFS, the BS has to vacate a channel and avoid it for 30 minutes if radar activity is detected.

Table \ref{tab:comparison} provides a taxonomy of some of the regulations for these bands in certain regions. Next, we discuss ALOHA-based and sensing-based access. The former primarily uses the sub-1GHz, whereas the latter enable IoT networks to use the bands at 2.4GHz and 5GHz.

\begin{table*}[!t]
\caption{Properties and regulations of different unlicensed bands}
\label{tab:comparison}
\scriptsize
\centering
\def\arraystretch{0.9}
\begin{threeparttable}[b]
\begin{tabular}{|c|c|c|c|c|c|c|c|}
\hline
Spectrum   				 &Bands  	  		& Regions\tnote{1}  	&Duty cycle & Peak power\tnote{2}        &Polite access &Comments &IoT networks\\\hline
\multirow{4}{*}{Sub-1GHz}&433-434MHz		&Europe		&$<10\%$	&10dBm EIRP			&No			   &BW limit may apply			&\multirow{4}{*}{\shortstack[c]{LoRa, Sigfox, \\Telensa, WavIoT\\ Weightless,\\ IEEE 802.11ah}}\\\cline{2-7}
						 &779-787MHz		&China		&$<0.1\%$	&12dBm EIRP			&No			   &--			&\\\cline{2-7}
						 &863-870MHz		&Europe		&$0.1-10\%$&14-27dBm	EIRP	&Optional LBT\tnote{3}  &Divided into subbands\tnote{4} 			&\\\cline{2-7}
					 	 &902-928MHz		&US			&No			&24-30dBm	&No		&Dwell time limit may apply				    &\\\hhline{|=|=|=|=|=|=|=|=|}
\multirow{2}{*}{2.4GHz\tnote{5}}  &\multirow{2}{*}{2402-2484MHz}	&Europe &No &20dBm 	    &LBT	&--					&\multirow{2}{*}{\shortstack[c]{Ingenu\\ Wi-SUN}}\\\cline{3-7}
						 &					&US					&No &30dBm		&No		&Dwell time limit may apply	&\\\hhline{|=|=|=|=|=|=|=|=|} 

\multirow{9}{*}{5.0GHz\tnote{6}}&
\multirow{3}{*}{5150-5350MHz\tnote{7}}	
&Europe		&No	&23dBm EIRP			&LBT/DFS			   	&Indoors only			&\multirow{9}{*}{\shortstack[c]{Cellular variants \\that support \\LTE-M and \\ NB-IoT\\such as \\MulteFire}}\\\cline{3-7}
&&US		&No	&24-30dBm 			&DFS			   		&DFS in 5250-5350MHz&\\\cline{3-7}
&&China		&No	&23dBm EIRP			&DFS			   		&Indoors only		&\\\cline{2-7}
&\multirow{3}{*}{5470-5725MHz}	
&Europe		&No	&30dBm EIRP			&LBT/DFS			&--			&\\\cline{3-7}
&&US		&No	&24dBm 				&DFS			   	&--			&\\\cline{3-7}
&&Japan		&No	&Depends on BW		&LBT/DFS			&--			&\\\cline{2-7}
&\multirow{3}{*}{5725-5875MHz}
&Europe		&No	&33dBm EIRP			&DFS			   	&Fixed wireless service only &\\\cline{3-7}
&&US		&No	&30dBm 				&No			   		&--			&\\\cline{3-7}
&&China		&No	&33dBm EIRP			&DFS			   	&--		&\\\hline
\end{tabular}
\begin{tablenotes}
	\item[1] These are sample regions. In Europe, rules can be found in ETSI documents, whereas they can be found in the FCC Title 47 Part 15 for the US. 
	\item[2] Some regulations are defined based on the peak transmit power, whereas others are based on the EIRP, which typically includes antenna gains. 
	\item[3] The presence of LBT affects the duty cycle limit.
	\item[4] Different subbands have different duty cycles, peak transmit powers, and additional regulations on the signal bandwidth (BW). 
	\item[5] The 2.4GHz is global, making regulations very similar across countries and regions. 
	\item[6] There are other bands for considerations, e.g., 5350-5470MHz and the DSRC band 5850-5925GHz. In China,  5470-5725MHz is under consideration. 
	\item[7] This can be further divided into 5150-5250MHz and 5250-5350MHz as regulations slightly differ between them in some regions.
\end{tablenotes}
\end{threeparttable}
\end{table*}

\section{Spectrum sharing using ALOHA-based access}\label{sec:ALOHA}
In ALOHA-based networks, access is random, i.e., the IoT device sends a packet at anytime, with restrictions primarily on the duty cycle or dwell time. Several IoT networks have opted for this paradigm for the following reasons. First, it reduces the control overhead, which is critical given that the IoT payload is generally small. Second, it relaxes synchronizing devices with the network, which reduces the complexity, cost, and energy consumption of the IoT device. 

To address the intra- and inter-network sharing issues, ALOHA-based networks universally use narrowband communications \cite{Yang2017a}. For example, networks that use UNB communications include Sigfox (100Hz-600Hz) and Telensa (500Hz). Further, networks like LoRa, although use CSS, are still considered narrowband (125KHz-500KHz). Such narrowband signaling brings two benefits. First, it divides the band into many channels, connecting more devices even in the spectrum-limited sub-1GHz bands. Second, for the same receiver sensitivity, the received signal-to-noise ratio (SNR) is improved as the noise bandwidth is much lower for narrowband channels, e.g., the thermal noise power in a 100Hz channel is 53dB lower than that in a 20MHz channel. The SNR gain can be utilized to enhance the IoT network's spectral efficiency or coverage \cite{Yang2017a}. 

Although narrowband communications has become central to ALOHA-based networks, the intra-network interference due to the asynchronous access can be significant for high densities of IoT devices. In addition, many of these networks have limited or no feedback (ACKs/NACKs). Because of these reasons, diversity techniques generally complement narrowband communications, as we discuss next. 

\subsection{The role of geographical diversity}
\begin{figure}[t!]
	\center
	\includegraphics[width=3.5in]{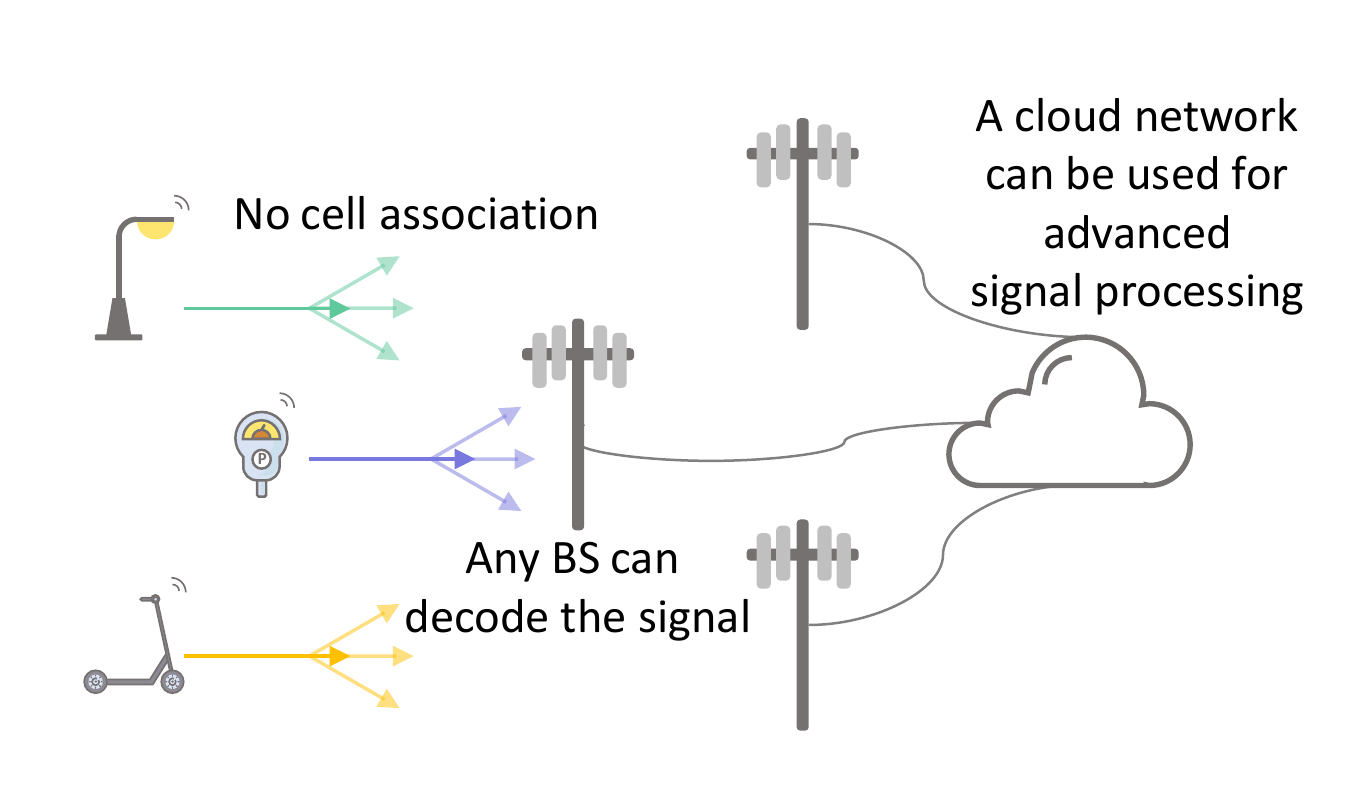}
	\caption{An illustration of the star architecture.}
	\label{fig:star}
\end{figure}

In geographical diversity, also known as macro diversity, multiple physically apart BSs are used to receive the same signal. Thus, the signal can experience multiple independently faded channels, substantially improving the transmission reliability. We distinguish this diversity method from spatial diversity, which is used by having multiple antennas at the same BS. The latter is more common in human-centric wireless networks, as the device typically associates with a single BS prior to data communications. However, in several IoT networks, such association phase is dropped to reduce control overhead. Thus, the device sends its data packet, without appending it with a cell ID, expecting at least one BS in vicinity can decode it, i.e., the IoT device in essence operates in an uplink broadcast mode. Figure \ref{fig:star} shows a typical star architecture of ALOHA-based IoT networks, which is used by LoRa and Sigfox. In such architecture, a BS can decode the signal, forwarding the output to the cloud, or it can pass the received signal to the cloud for more advanced processing. The cloud also handles any potential replicas of the same packet, so that only one successfully decoded packet is stored or sent to the application layer. 

In \cite{HattabCabric2018b}, an analytical framework is developed to model massive IoT networks in the presence of interfering networks. The framework is used to study the impact of geographical diversity on the transmission success probability and the IoT connection density. It is shown that when the device is not restricted to communicate only with its nearest BS, the network operator can significantly reduce the density of BSs needed to cover the network. Alternatively, for the same density of BSs, no BS association tangibly increases the connection density of the network.

\subsection{The role of repetition diversity}
Repetition diversity can be used to complement geographical diversity, where the same packet is sent multiple times over time and/or frequency. For example, Sigfox typically uses three transmissions per message, whereas LoRa has a reconfigurable number of transmissions from 1 to 15. Repetition diversity is also adopted for cellular-based solutions such as NB-IoT. 

Due to the asynchronous nature of access and the absence of cooperation between BSs, each BS tries to independently decode each transmission. Thus, a BS can decode any of these repeated transmissions only if the transmission with the maximum SINR is above the decoding threshold. Indeed, it is shown in \cite{HattabCabric2018b} that if the IoT device sends the packet $N$ times, a processing gain of $\sum_{n=1}^N\frac{1}{n}$ is achieved, which is identical to the well-known processing gain of selection diversity. Nevertheless, sending the packet multiple times increases power consumption for a given payload (and latency, yet it is typically less critical), and it worsens the intra-network interference due to congesting the network with more transmissions. In fact, it is shown in \cite{HattabCabric2018b} that a single transmission maximizes the transmission success probability when the density of IoT devices is much higher than that of incumbents.

\subsection{Further enhancements for ALOHA-based IoT networks}
In addition to the aforementioned diversity schemes, sophisticated signal processing can be used to combine packet replicas at a single BS, e.g., coherent combining,  or across BSs, e.g., distributed MIMO. The latter for example is implemented in  \cite{Dongare2018}. In particular, when a LoRa BS detects IoT packets, it forwards the collected I/Q samples to the cloud. The cloud, which collects I/Q samples from multiple BSs, corrects for any timing and frequency offsets. Afterwards, it calculates a feature vector for each packet, taking into account the packet reception time and the location of the BS that forwarded it. The feature vector is used to cluster packets, where coherent combining within each cluster is later performed. Finally, the combined packets from each cluster are jointly decoded. This proposed system is shown to outperform the standard LoRa scheme in terms of coverage.

Alternative to the diversity schemes, densifying the network with BSs or increasing the number of bands can further improve the connection density. The former incurs additional capital expenditure on the IoT network operator, whereas the latter requires complex hardware. Indeed, a Sigfox BS listens to a single multiplexing band of bandwidth $B=200$KHz, where IoT devices communicate using UNB signals, with center frequencies uniformly selected within this band. Because there are no clear subchannel boundaries within the band, the BS samples the spectrum at a very fine resolution, e.g., the fast Fourier transform size for $B=200$KHz is at least $2^{14}$ \cite{HattabCabric2018b}. Hence, using wider multiplexing bands can significantly increase the computational complexity at the Sigfox BS. In \cite{HattabCabric2018b}, we show that if we increase $B$ to $M\cdot B$, yet restrict each BS to randomly select one of the $M$ multiplexing bands, then the IoT network can still gain in terms of coverage and connection density, although the number of BSs listening to a given band is reduced by a factor of $1/M$. To achieve this, however, the device must send each transmission at a different band. Otherwise, if a device sends all packets at one randomly selected band, the performance does not improve compared to single-band IoT networks, i.e., the gain achieved by having fewer collisions within a band is completely canceled by the loss of having fewer BSs listening to a given band. In other words, sending packets at different bands attains additional geographical diversity as each packet is received by a different subset of BSs.

\subsection{Simulation Results}
In this section, we study how the aforementioned diversity schemes affect ALOHA-based access in terms of the transmission success probability\footnote{The success probability can also be interpreted as the complementary cumulative distribution function (CCDF) of the received signal-to-interference-plus-noise ratio (SINR).} and the network connection density, i.e., how many devices can be connected for a given success probability constraint. 

We use the framework in \cite{HattabCabric2018b} and simulate a Sigfox network with US specifications, where IoT devices generate six packets per hour.  We randomly deploy Sigfox BSs and 30,000 Sigfox devices per BS in a given region. We assume that the spectrum is shared with LoRa, as an incumbent network, where we drop 1,000 LoRa devices per Sigfox BS, with each one using CSS of bandwidth 125KHz. All devices transmit at 14dBm. We compare between three variants of Sigfox networks: (i) nearest BS association, (ii) existing, i.e., no BS association, and (iii) the proposed multiband access with $M=5$, i.e., the network uses 1MHz of the spectrum.

\begin{figure*}[t!]
	\centering
	\begin{subfigure}[t]{.45\textwidth}
		\centering
		\includegraphics[width=3in]{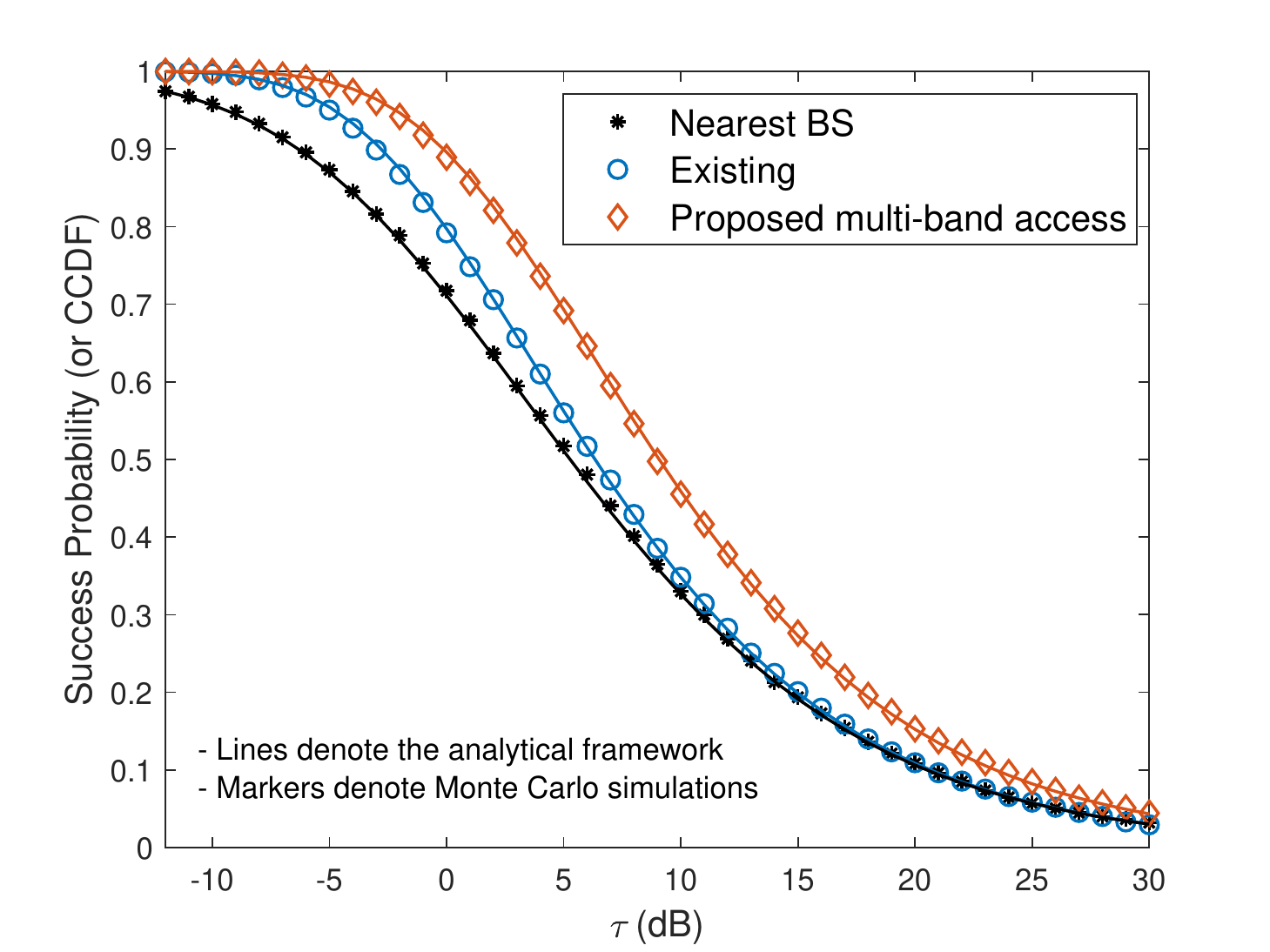}
		\caption{}
		\label{fig:Ps_CCDF}
	\end{subfigure}
	\begin{subfigure}[t]{0.45\textwidth}
		\centering
		\includegraphics[width=3in]{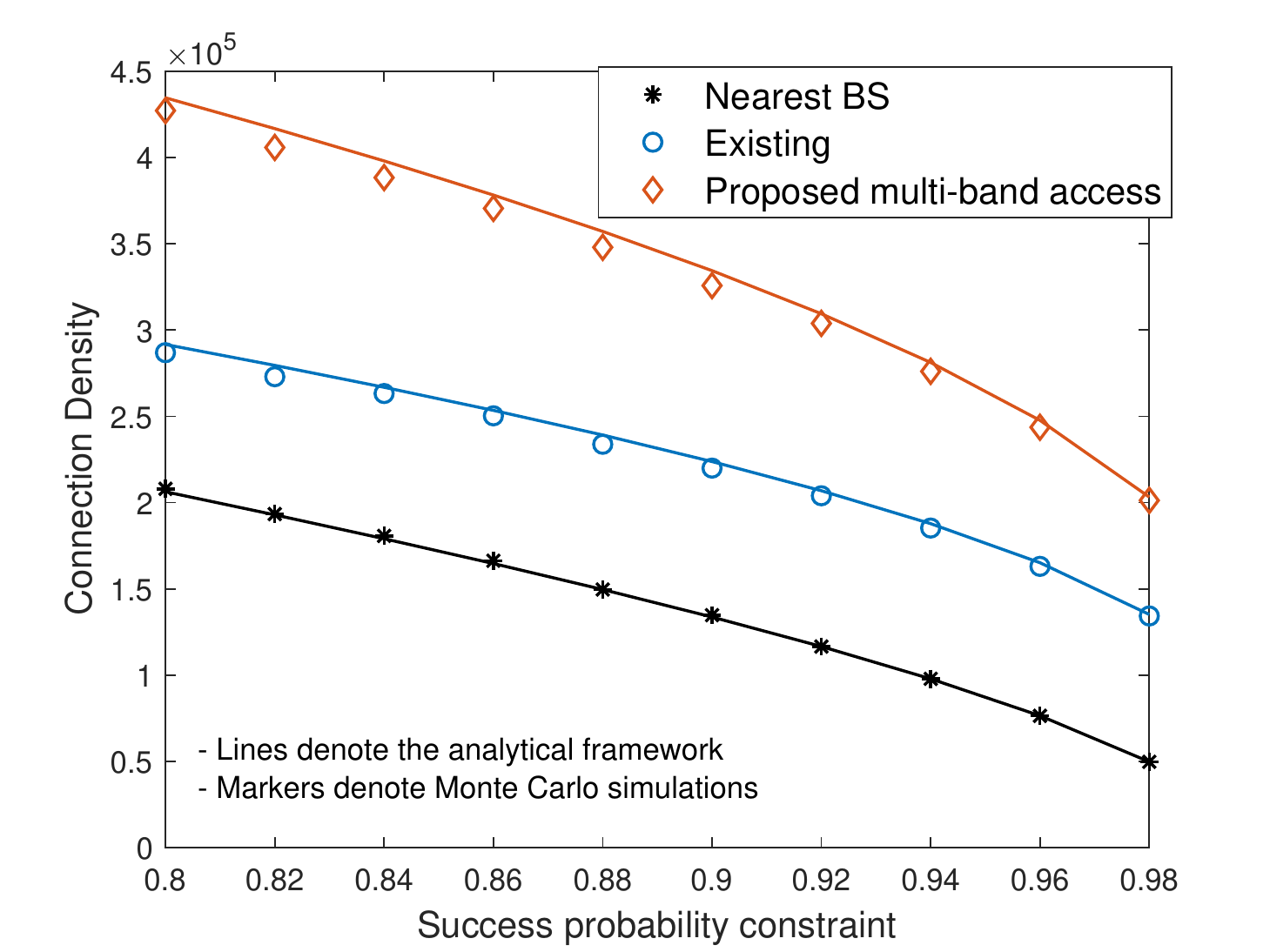}
		\caption{}
		\label{fig:connectionDensity}
	\end{subfigure}\\
	\begin{subfigure}[t]{.45\textwidth}
		\centering
		\includegraphics[width=3in]{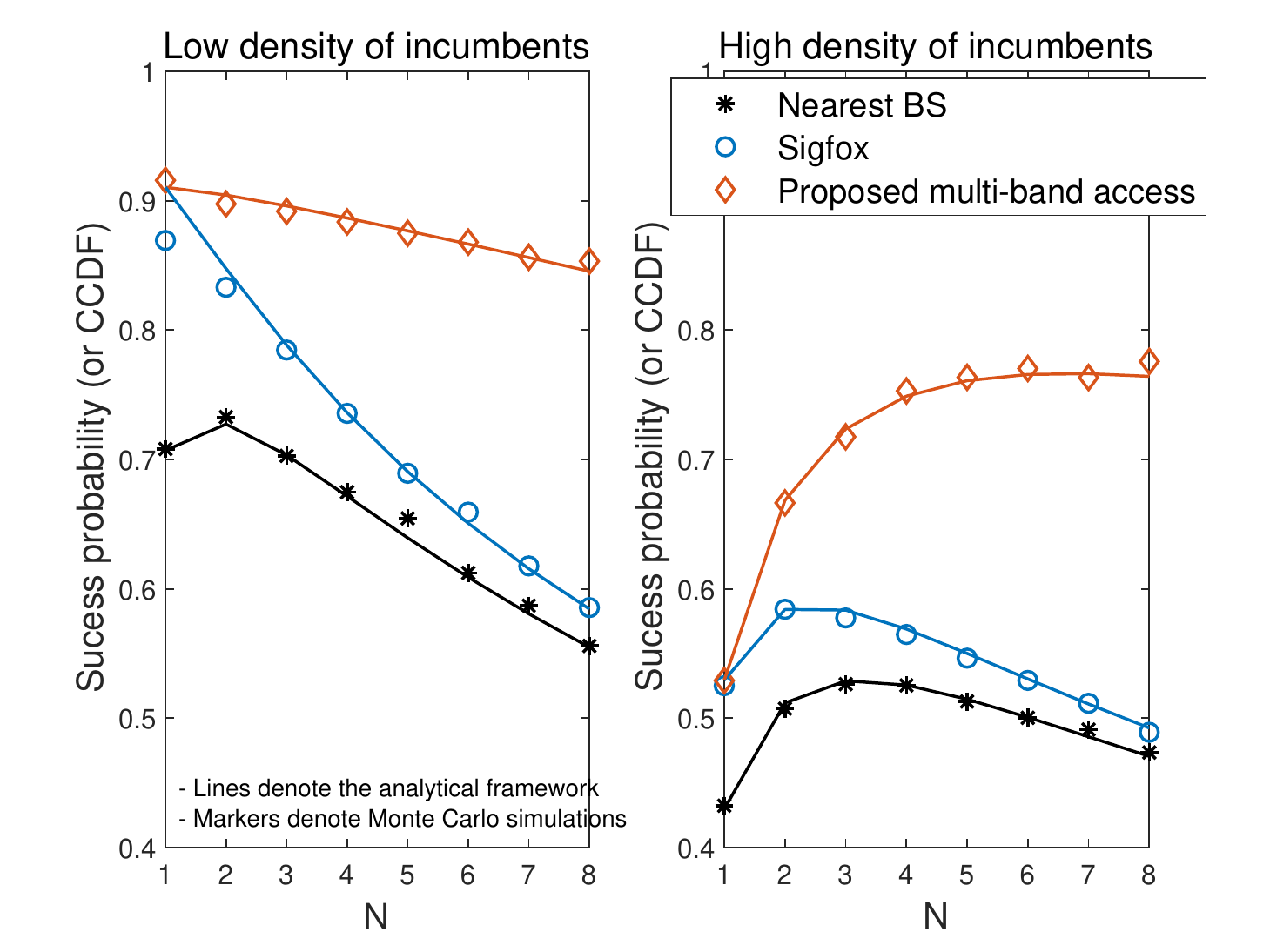}
		\caption{}
		\label{fig:Ps_N}
	\end{subfigure}
	\caption{The role of diversity in ALOHA-based networks: (a) Success probability performance for a given decoding threshold; (b) Connection density for a given success probability constraint ($\tau=0$dB); (c) Success probability for different number of packet transmissions ($\tau=0$dB).}
	\label{fig:ALOHAbased}
\end{figure*}

In Fig. \ref{fig:Ps_CCDF}, we show the transmission success probability of Sigfox devices for a given decoding threshold $\tau$. It is observed that the cell-edge SINR, i.e., the 95th percentile, is improved by 5dB when IoT devices do not append their packets with a cell ID, allowing any BS to decode the packet. Multiband access adds an additional 3dB gain compared to the existing Sigfox network. Such SINR improvements are translated into tangible gains in the connection density. Indeed, in Fig. \ref{fig:connectionDensity}, we show the connection density of the Sigfox network with 25 BSs randomly deployed in a 25x25km$^2$ area. With no BS association, the connection density improves significantly, e.g., for a 98 percent target success probability, Sigfox with geographical diversity increases the connection density by 2.7x compared to the network with nearest BS association, and it increases it by 4x when mutliband access is used. 

Figure \ref{fig:Ps_N} shows the success probability with variations of the number repetitions, $N$, when the density of incumbents is low (1,000/BS) and high (30,000/BS). It is clear that a single transmission is optimal for Sigfox with no BS association when the incumbents' density is low. As interference becomes dominated by the incumbent network, a higher value of $N$ can help improve the network performance, yet sending too many repetitions eventually degrades the performance. Another interesting observation is that more repetitions would be needed if IoT devices are restricted to connect to a single BS, i.e., higher repetition diversity is required when geographical diversity is absent. Finally, multiband access is shown to be less sensitive to intra-network interference compared to single-band access, as allowing the device to have a larger pool of channels to select from reduces collisions.

\section{Spectrum sharing using sensing-based access}\label{sec:sensing}
ALOHA-based IoT networks are suitable for very low-rate applications (bps to kbps). For applications that have higher rate requirements (few Mbps), wider bandwidths are needed. Even if wide bands are available in the limited sub-1GHz spectrum, the received SNR decreases for a given receiver sensitivity when IoT devices use wider channels, which can be detrimental to networks that rely on random access.

The higher unlicensed bands, e.g., 2.4GHz and 5GHz, provide opportunities for massive IoT applications with more varying rate requirements. These bands have been primarily used by WiFi networks, yet ad-hoc networks are not suitable for wide-area coverage. Cellular networks, on the other hand, are ubiquitous and its existing infrastructure can be leveraged to provide wide-area connectivity. More importantly, two fronts of ongoing cellular standardization can accelerate the adoption of massive IoT over these bands. The first one is the use of unlicensed access over cellular networks and the second one is the newly introduced user categories tailored for low-cost IoT devices. For example, MulteFire is a 3GPP-compliant standard that enables LTE to operate at 2.4GHz and 5GHz in a standalone manner, i.e., without any licensed carriers \cite{MulteFire2017}. In addition, LTE-M and NB-IoT use 1.4MHz and 180KHz channels, respectively, which help reduce the device complexity and divide a 20MHz channel into many resource blocks that can be allocated to a larger number of IoT devices \cite{Wang2017b}. 

The key difference of using the higher bands compared to the sub-1GHz band is the need for sensing-based access, e.g., LBT. While such requirement is specific to certain regions, the desire to have a globally harmonized network have pushed 3GPP standardization to adopt LBT.

\subsection{Non-cooperative LBT networks} 
A key market for MulteFire is private IoT networks, which can be either operated by MNOs or neutral hosts \cite{MulteFire2017}. MulteFire brings the benefits of cellular-based access, e.g., synchronization, coordinated access, and centralized resource allocation,  at no additional cost of licensing a band. Each MulteFire BS implements LBT, using energy-based sensing to determine whether to access the channel. We refer to this mechanism as narrowband non-cooperative LBT, as BSs do not coordinate sensing or access among them.

A key limitation to the current LBT mechanism is that only one band, of bandwidth 20MHz, is sensed, and so the MulteFire network cannot handle massive IoT networks. To scale MulteFire, the network must be able to identify a large number of narrowband channels in a wideband spectrum and to aggressively reuse them over space. The former requires sensing at a fine spectral resolution, e.g., sensing a spectrum of 500MHz at 180KHz resolution. The latter requires capturing the footprints' of active incumbents at a fine spatial resolution, e.g., reusing the same channels across small cells that are 100-200m apart. 

\subsection{Distributed wideband LBT networks}
If the 5GHz spectrum, with approximately 500MHz bandwidth, is divided into channels of bandwidth 180KHz for NB-IoT operation, then each MulteFire BS needs a spectrum scanner that can quickly sense 2775 channels! Equipping a dense network with such complex scanners may incur prohibitive costs. To this end, we have proposed in \cite{Hattab2018a} a sensing assignment scheduler that divides the sensing tasks among BSs under the following constraints: each BS senses a single band of bandwidth 20MHz, i.e., 111 NB-IoT channels, and each band within the 500MHz spectrum is sensed by at least a certain number of BSs. The scheduler aims to ensure that the distance between a BS and any BSs sensing bands not sensed by that BS is minimized. 

By ensuring that each band is sensed by multiple BSs, cooperation among BSs can be used to further enhance spectrum sensing reliability and combat the hidden terminal problem, particularly because energy-based sensing is generally unreliable in fading channels. Nevertheless, using a centralized processing of sensing reports collected by the cooperating BSs destroys spatial information about incumbents’ energy footprints, i.e., all cooperating BSs arrive at the same sensing decision, limiting channel reuse across space. To mitigate this issue, a distributed sensing algorithm is developed in \cite{Hattab2018a}. In the distributed algorithm, each BS exchanges few reports only with its neighboring BSs, limiting the communication overhead. For each message exchange, the BS uses an adaptive filter, combining the collected reports with different weighting coefficients. The coefficients are used to help quickly diffuse spectrum information across the network.  At the end of the distributed sensing stage, each BS acquires local knowledge about the wideband spectrum even though only a small portion of the spectrum is sensed by each BS. 

\subsection{Case study: Massive IoT in public parks} 
\begin{figure*}[t!]
	\centering
	\begin{subfigure}[t]{.45\textwidth}
		\centering
		\includegraphics[width=3in]{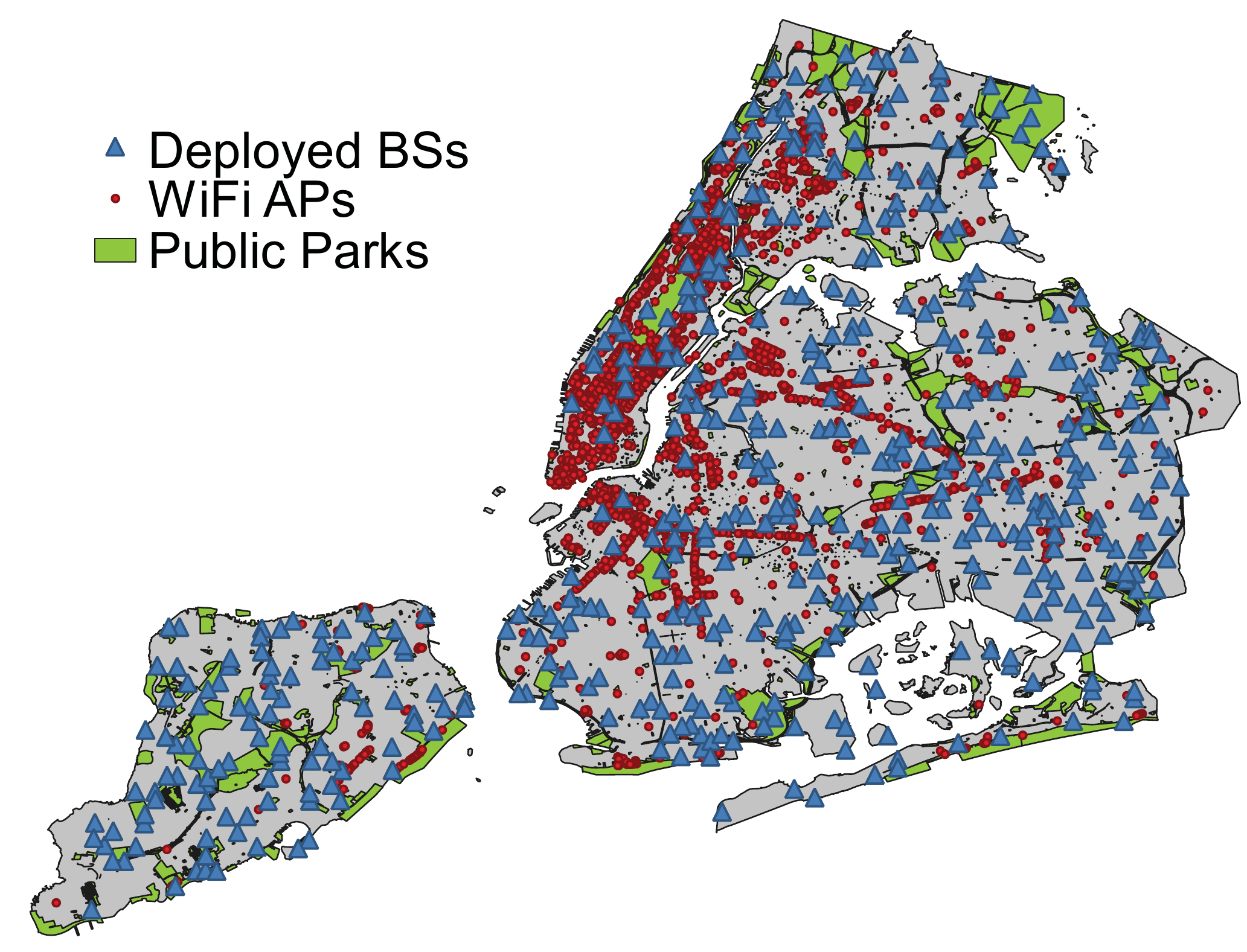}
		\caption{}
		\label{fig:NYC}
	\end{subfigure}
	\begin{subfigure}[t]{.45\textwidth}
		\centering
		\includegraphics[width=3in]{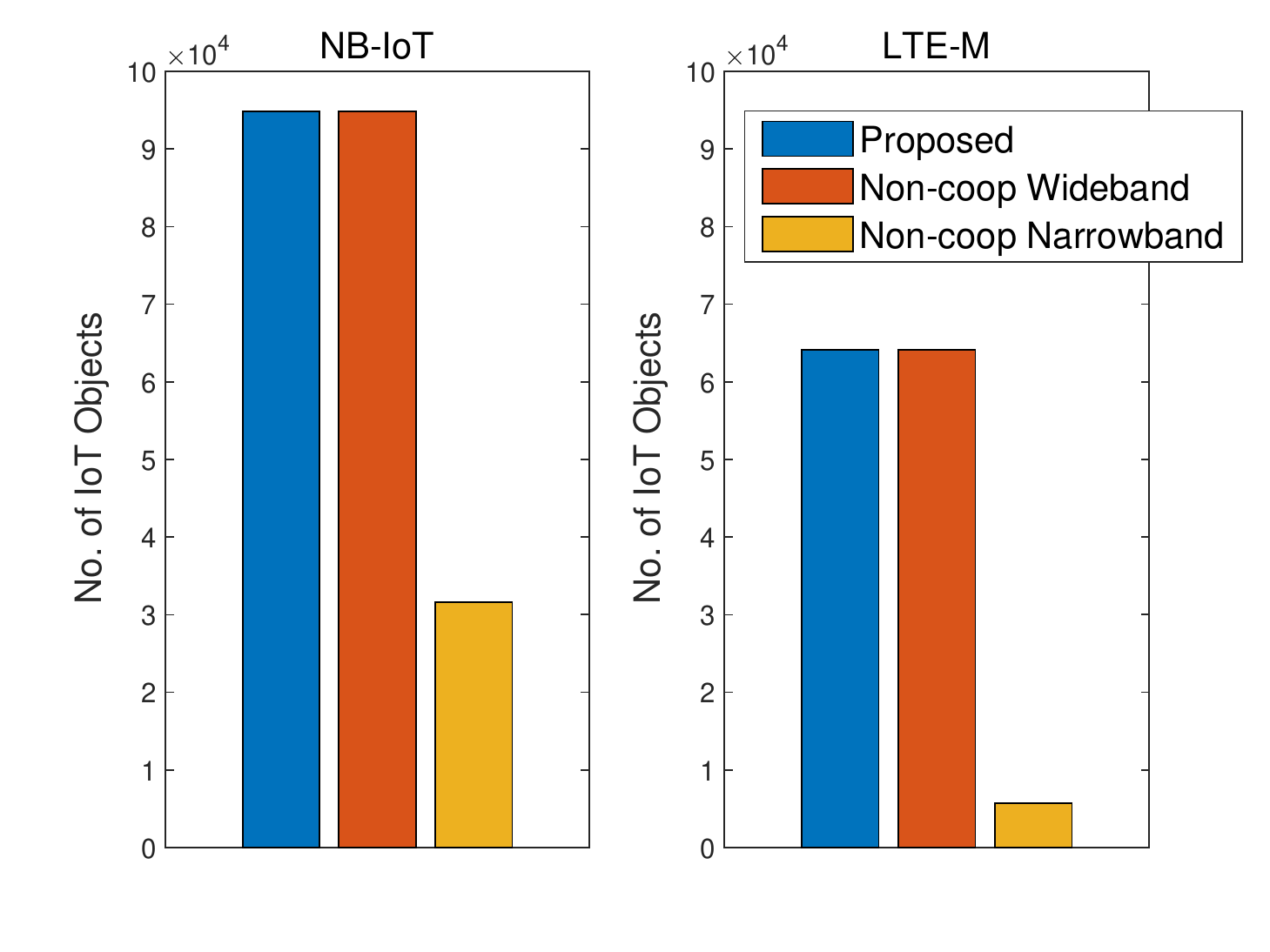}
		\caption{}
		\label{fig:CaseStudySim}
	\end{subfigure}
	\caption{Case study for massive IoT in public parks: (a) An illustration of the simulation set-up; (b) Average number of scheduled devices per scheme.}
\end{figure*}

We simulate a large-scale deployment of $10^5$ IoT sensors in the public parks of New York City (NYC), where such sensors can be used to record traffic activities in the park, monitor air quality, and meter water supply. We randomly drop 500 BSs across the city and use the NYC Open Data to extract the locations of 2000 actual outdoor public WiFi access points (APs) in NYC, treating them as interfering incumbents. The APs are assumed to transmit at 30dBm, and each occupies either 20MHz, 40MHz, or 80MHz within a wideband spectrum of bandwidth 500MHz centered at 5.4GHz. An illustration of the set-up is shown in Fig. \ref{fig:NYC}. 

We compare between three MulteFire networks: (i) non-cooperative narrowband, where each BS randomly selects a 20MHz band to sense, (ii) non-cooperative wideband, where each BS senses the entire spectrum, and (iii) the proposed distributed wideband architecture, where a sensing assignment scheduler assigns each BS 20MHz to sense, and then each BS uses the distributed sensing algorithm to learn the occupancy of the entire spectrum. A channel is identified as available if its measured energy is below $-62$dBm.

Figure \ref{fig:CaseStudySim} shows the average number of devices that are scheduled over channels that are correctly identified as available for each scheme. It is observed that for NB-IoT, almost all devices are scheduled via the proposed and the non-cooperative wideband schemes, yet the latter requires wideband spectrum scanners at each BS. Equally important, compared to the non-cooperative narrowband scheme, the proposed system increases the connection density by 3x and 11x for NB-IoT and LTE-M operations, respectively.

\section{Future Research Directions}
ALOHA-based and sensing-based access remain the dominating spectrum sharing methods for massive IoT, yet there are alternative options that need further exploration such as database-assisted access, spectrum access systems, and drone-assisted access.

\subsection{Database-assisted access}
The TV white space (TVWS) has been recently utilized to provide broadband connectivity, particularly in rural areas. For example, there is approximately 180MHz of spectrum (512-698MHz) that can be used by IoT networks given that they do not interfere with incumbents such as TV stations. To this end, a geolocation database is used to identify which channels can be used. One example of using TVWS is Microsoft's FarmBeats project, where IoT sensors collect farm-related physical parameters, e.g., soil moisture, and these measurements are forwarded to the cloud using the TVWS. 

In addition to the TVWS, the authors in \cite{Khan2017} explore the use of radar bands. In particular, a zone-based sharing access framework is proposed, where a repository containing the radars' locations and rotation rate is used to identify available bands. 

\subsection{Spectrum access systems}
In addition to licensed and unlicensed access, a third emerging paradigm is the licensed-shared access (LSA) or spectrum access systems (SAS), which comprises different tiers of users with varying access privileges. For example, the FCC has freed up 150MHz in the 3.5GHz CBRS band and introduced three tiers of access: incumbents access, priority access license (PAL), and general authorized access (GAA). PAL users can acquire spectrum via an auction, whereas GAA users do not need a license. However, access priorities are given to incumbents followed by PAL users, e.g., if a PAL user is active over a channel used by the GAA user, the latter must vacate the channel immediately. 

Acquiring a PAL adds additional costs on IoT network operators, and thus GAA may be desirable to pursue instead, yet a dynamic spectrum access mechanism would be needed to ensure that incumbents and PAL users are protected from any interference.

\subsection{Drone-assisted access}
The use of drones as data aggregators or BSs has become an emerging trend for future connectivity solutions. The mobility of drones brings unparalleled flexibility to the wireless network. Indeed, drones can be used to extend the coverage of existing infrastructure and is more robust to natural disasters that could affect the network's infrastructures. For IoT networks, drones can reduce the communication range to IoT devices, which allows IoT devices to transmit at low power, extending their lifetime. 

The authors in \cite{Motlagh2017} have already developed an IoT platform that uses drones for crowd surveillance, showing the feasibility of using drones as data aggregators. Nevertheless, scaling such platform to cover wide areas still requires innovative solutions to fly drones for extended durations and to autonomously control a network of drones to perform specific tasks.

Parallel to the aforementioned research directions, regulatory and standard bodies are exploring different parts of the spectrum for future unlicensed-based access. A primary example is the FCC's notice of proposed rulemaking (FCC 18-147) regarding expanding unlicensed-based access to the 5.9-7.1GHz bands. For example, the subbands in 6.425-6.525GHz and 6.875-7.125GHz are proposed for low-power indoor operations, whereas the subbands in 5.925-6.425GHz and 6.525-6.875GHz can be used for outdoor applications, yet an \emph{automated frequency control} (AFC) system is required to protect incumbents such as fixed-satellite service. While the FCC is currently seeking more comments on several aspects of the AFC system, it is currently envisioned to be a simple database, for which BSs can request a list of available frequencies at their locations. For this reason, future research can focus on the development of AFC systems that consider the traffic properties of wide-area IoT networks.

\section{Conclusion}\label{sec:conclusion}
Enabling massive IoT connectivity over the unlicensed spectrum requires innovative spectrum sharing methods to connect a high density of IoT devices and to harmoniously coexist with incumbents. To this end, narrowband communications and geographical diversity are paramount for ALOHA-based access. For sensing-based access, it is critical to identify spectral opportunities at a fine spectral and spatial resolution, elevating the need for low-cost distributed wideband sensing algorithms.

\bibliographystyle{IEEEtran}
\scriptsize\bibliography{C:/Users/ghait/Dropbox/References/IEEEabrv,C:/Users/ghait/Dropbox/References/References}

\end{document}